\documentstyle[12pt,epsfig]{article}
\newcommand{\blankline}{\vskip .3cm}
\newcommand{\f}{\begin{equation}}
\newcommand{\ff}{\end{equation}}
\renewcommand{\thefootnote}{\fnsymbol{footnote}}
\def\be{\begin{equation}}
\def\ee{\end{equation}}

\def\ea{\end{eqnarray}}

\def\SLAC{Stanford Linear Accelerator Center and ITP\\
    Stanford University, Stanford, California 94309 USA}
\def\doeack{\footnote{Work supported by the Department of Energy,
                     contract DE--AC03--76SF00515.}}
\def\PI{Perimeter Institute for Theoretical Physics, Waterloo, Canada }

\def\RIT{Rochester Institute of Technology\\
84 Lomb Memorial, Rochester, New York, 14623-5604, U.S.A}

\def\Title#1{\begin{center} {\Large #1 } \end{center}}
\def\Author#1{\begin{center}{ \sc #1} \end{center}}
\def\Address#1{\begin{center}{ \it #1} \end{center}}
\def\andauth{\begin{center}{and} \end{center}}
\def\submit#1{\begin{center}Submitted to {\sl #1} \end{center}}

\def\submit#1{\begin{center}Submitted to {\sl #1} \end{center}}

\begin{document}
\begin{titlepage}

\vfill
\Title{The Gravitational Instability of the Vacuum: Insight into the Cosmological Constant Problem}
\vfill
\Author{Stephon Alexander\doeack}
\Address{\SLAC}
\medskip
\andauth
\medskip
\Author{Manasse Mbonye}
\Address{\RIT}
\medskip
\andauth
\medskip
\Author{John Moffat}
\Address{\PI}
\vfill

\vfill
\submit{Physical Review D}
\vfill
\end{titlepage}
\def\thefootnote{\fnsymbol{footnote}}
\setcounter{footnote}{0}
\newpage

  \centerline{ABSTRACT}
A mechanism for suppressing the cosmological constant is
developed, based on an analogy with a superconducting phaseshift
in which free fermions coupled perturbatively to a weak
gravitational field are in an unstable false vacuum state. The
coupling of the fermions to the gravitational field generates
fermion condensates with zero momentum and a phase transition
induces a nonperturbative transition to a true vacuum state by
producing a positive energy gap $\Delta$ in the vacuum energy,
identified with $\sqrt{\Lambda}$, where $\Lambda$ is the
cosmological constant. In the strong coupling limit a large
cosmological constant induces a period of inflation in the early
universe, followed by a weak coupling limit in which
$\sqrt{\Lambda}$ vanishes exponentially fast as the universe
expands due to the dependence of the energy gap on the density of
Fermi surface fermions, $D({\epsilon})$, predicting a small
cosmological constant in the present universe.

 \blankline\blankline\blankline\blankline\blankline
 \blankline\blankline\blankline\blankline\blankline\blankline
 \blankline
emails:  salexander@itp.stanford.edu, mrmsps@rit.edu, jmoffat@perimeterinstitute.ca
 \eject
\tableofcontents
\newpage

\section{Introduction}

The cosmological constant problem centers around two questions:

1) Why is the cosmological constant observed to be much smaller than the expected Planckian
value?

2)  If we choose the bare cosmological constant to be the observed
value today, what is the mechanism which stabilizes it under
quantum corrections~\cite{Weinberg,Straumann}?

Many investigators have proposed solutions ranging from a rolling
scalar field to the anthropic
principle\cite{Ratra,Susskind,Caldwell,Alexander}¥. Moreover, it can be
argued that the greatest irony of the cosmological constant
problem is the inflationary
paradigm\cite{Brandenberger,Mbonye,Woodard}. The onset of the
inflationary epoch is completely dominated with a fluid that is
either a pure cosmological constant or a scalar field whose
potential is extremely flat, mimicking a cosmological constant.
Clearly this period of inflation has to end for successful
structure formation. However, whatever mechanism is responsible
for this exit from inflation should be a clue as to how the
cosmological constant is relaxed at all times including today.

In the following, we shall propose a mechanism to solve the
cosmological constant problem, based on a simple idea analogous to
the microscopic realization of superconductivity. We argue that
the perturbative vacuum state in the presence or absence of a
cosmological constant is gravitationally unstable and it is
energetically favorable for the vacuum associated with the
effective cosmological constant to release all of its energy into
the production of condensates, bound states of the free fermions.
The formation of condensates leads to a non-perturbative true
ground state.  Similarly, Bose-Einstein condensates have been
proposed by \cite{zhit,Ball,Parker,Arkani-Hamed,Silverman} as a viable alternative for a
fundamental inflaton scalar field, because they can have a nonzero
potential sufficiently flat to lead to inflation. Condensates also
enjoy the property of ending inflation since they can vanish in
the infra-red (at late times).

In the absence of a gravitational interaction between fermions the
Minkowski flat spacetime has a zero cosmological constant. When
the gravitational interaction is switched on, the Minkowski
spacetime vacuum becomes unstable, and the universe enters into a
superfluid phase of fermion condensates. Below a critical phase
transition temperature, $T_c$, the binding energy of a pair of
fermions causes the opening of a positive energy gap $\Delta$ in the
ground state (vacuum energy) of the fermion condensate system. The
positive energy gap $\Delta$ is identified with the square root of
the cosmological constant, $\sqrt{\Lambda}$. We shall describe the
formation of the non-perturbative energy gap due to the exchange
of gravitons between fermions, in analogy with the exchange of
phonons between electrons in a crystal structure in a
non-relativistic Bardeen, Cooper and Schrieffer
model~\cite{Bardeen}. In an initial strong coupling limit the
energy gap $\Delta$ or $\sqrt{\Lambda}$ yields a large enough
cosmological constant to induce a period of de Sitter inflation.
This is followed by a weak coupling limit as the universe accelerates,
leading to an exponential suppression of $\Lambda$ and a
``graceful'' exit to inflation.

The non-perturbative nature of the vacuum instability and the
formation of a vacuum energy gap in the early universe, explains
why any attempt to derive the cosmological constant from a {\it
perturbative} quantum field theory calculation leads to an
egregious disagreement with the observed value of the vacuum
energy, when the latter is identified with dark energy.

This paper is organized as follows:  Section II begins with the
construction of a Fermi liquid in de Sitter Space.  In section
III, we consider a non-relativistic derivation of the energy gap
displaying both the weak and strong coupling limit of the
fermionic condensate and its relation to the exponential
suppression of the cosmological constant. In section IV, we
discuss a relativistic field theory model of the formation of
$\Lambda$ in the vacuum energy due to gravity and a screened
fermion attractive force, based on the
Nambu-Jona-Lasinio~\cite{Nambu} 4-fermion interaction model. We
conclude the paper, in section V, with a summary of the results
and a discussion of future directions of research.

\section{The de Sitter Fermi Liquid}

In this section, we want to draw some similarities between
fermions in de Sitter space and fermions in a superconductor. This
analogy was made precise in the Nambu-Jona-Lasinio~\cite{Nambu,Inagaki}
model of the four fermion interaction. Our main point is to derive the density of states
from fermions which naturally arise in de Sitter space and derive its dependence on the scale factor.  We now
proceed to construct a Fermi-surface in de-Sitter space from
N-fermions.

Let us recall that de Sitter space is the maximally symmetric
solution of the Einstein equations with a positive cosmological
constant.  It is generated by an isometry group of $SO(4,1)$ and
can be seen as a timelike hyperboloid embedded in a $4+1$
dimensional Minkowski space obeying the constraint: \be R^{2} =
-T^{2} + X_{1}^{2} + X_{2}^{2} + X_{3}^{2} + X_{4}^{2} \ee One can
consider a special class of geodesic observers in de Sitter space
which corresponds to a fermionic representation.  For example, by
specifying the observer at the north pole in the positive $X_{4}$
direction, one is left with Lorentz generators which leave the
observer's worldline invariant under rotations about the axis
connecting the poles as well as the boosts in the $X_{4}$
direction.  These generators form the observer's little group
which is $S0(3)\times R$, whose representations correspond to
massless fermions. In general the generators of de Sitter can be
written as \be M_{IJ} = -\frac{i}{4}[\Gamma_{I},\Gamma_{J}] \ee ,
where the gamma matrices obey the Clifford algebra
$\{\Gamma_{I},\Gamma_{J}\}=2\eta_{IJ}$ and $I,J =0..4$. Let us
express the de Sitter generators by indices $\mu\;, \nu$  which
run from 1 to 3. \be J_{\mu\nu} =M_{\mu\nu} \;\; P_{\mu}=M_{4\mu}
\;\; K_{\mu}=M_{0\mu} \;\; H=M_{04} \ee ,

   \be  J_{\mu}= \left(
   \begin{array}{cc}
   \sigma_{\mu} & 0 \\
   0 & \sigma_{\mu}
   \end{array}
   \right) \; \;
     P_{\mu}= \frac{i}{2}\left(
   \begin{array}{cc}
  0 & \sigma_{\mu} \\
    -\sigma_{\mu} & 0

   \end{array}
   \right)
   \ee
   \be  K_{\mu}= \frac{i}{2}\left(
   \begin{array}{cc}
   0 & \sigma_{\mu} \\
   \sigma_{\mu} & 0
   \end{array}
   \right) \; \;
     H= \frac{i}{2}\left(
   \begin{array}{cc}
  -1 & 0 \\
    0 & 1

   \end{array}
   \right)
   \ee

These generators will act on two component spinors and will form a
many particle Hilbert space.  We can construct the Lagrangian for
free massless fermions in de Sitter space.
 \be \cal{L}\rm = \sqrt{g}\left(R(e) + \bar{\psi}e^{\mu}_{b}\gamma^{b}(i\partial_{\mu}-\frac{1}{2}
 \omega_{\mu cd}J^{cd})\psi\right) \ee
where $\omega_{\mu cd}=
 e^{\nu}_{d}\left(e_{\nu c,\mu}-\Gamma^{\rho}_{\mu\nu}e_{\rho c}\right)$ are the spin connection
 coefficients.

The $FRW$ background is given by the metric \be
ds^{2}=a^{2}(\eta)(-d\eta^{2} + dx^{i}dx_{i})\ee where $\eta$ is
the conformal time, noting that for de Sitter space $a=\frac{1}{H\eta}$.  Choosing for the vierbein,
$e_{\mu b}=a\eta_{\mu b}$, we can solve for the spin connection:
 \be \omega_{\mu cd}=\left(\eta_{\mu c}\partial_{d} -\eta_{\mu d}\partial_{c}\right)\ln(a) \ee

Upon substitution into the above Lagrangian we obtain
\be \cal{L}\rm =\frac{1}{2}(a^{3/2}\bar{\psi})i\gamma^{\mu}\partial_{\mu}(a^{3/2}\psi)
\label{action} \ee
and varying with respect to $\bar{\psi}$, we get the equation of
motion

 \be i\gamma^{\mu}\partial_{\mu}(a^{3/2}\psi)=0 \ee
 or, more explicitly
 \label{eqs}
 \be \gamma^{0}[H\psi + \dot{\psi}] + \vec{\sigma}\cdot\vec{\nabla}\psi=0 \ee
 where $H={\dot{a}}/a$ is the Hubble parameter.
Therefore the solution of a massless Dirac particle propagating in
de-Sitter space describes plane waves with a dispersion relation
\be \omega^{2}_{p} =k^{2} \ee

The density of states will therefore grow in an expanding
universe, if they are quantized in a co-moving box. \footnote{We
thank Lenny Susskind and B.J Bjorken for clarifying this issue
with one of us (SA).} We are interested in the density of states
of these fermions in order to make contact with condensation in
the following sections. The density of states can directly be
obtained from the definition: \be D(k)=\frac{dN}{dk}\frac{dk}{dE}
\ee After some straightforward algebra we get \be D(k) \sim
a^3(t). \ee This result will be of relevance in the following
sections.

\section{Vacuum Instability, Fermion Condensates, the Gap Equation and
the Cosmological Constant}

The cosmological constant problem can be stated as follows. We
have the Einstein gravitational equations \be R_{\mu\nu}
-\frac{1}{2}g_{\mu\nu}R-\Lambda_{0}g_{\mu\nu}= 8\pi GT_{\mu\nu},
\ee where $\Lambda_0$ is the ``bare'' cosmological constant and
\be T_{\mu\nu}=T^M_{\mu\nu}+T^{\rm vac}_{\mu\nu} \ee Here,  \be
T^{\rm vac}_{\mu\nu}=\rho_{\rm vac}g_{\mu\nu}\ee This leads to the
definition of an effective cosmological constant \be \Lambda_{\rm
eff}=\Lambda_0+8\pi G\rho_{\rm vac}\ee where $\rho_{\rm vac}$ is
the vacuum energy density. A calculation of the vacuum density for
a cutoff of order the Planck energy leads to a result that is 120
orders of magnitude larger than the observed
value~\cite{Weinberg,Straumann}.

Let us now consider a {\it non-perturbative} model to solve the
cosmological constant problem, based on an analogy with the
microphysical realization of superconductivity. We argue that in
the absence of gravitational interactions between fermions,
Minkowski spacetime is unstable and the cosmological constant
$\Lambda=0$. A {\it non-perturbative} phase transition to a true
vacuum state occurs when the gravitational interaction is taken
into account. The fermions form Cooper pair condensates with zero
momentum due to the weak gravitational interaction and a screened
long-range attractive interaction among the pairs of fermions. The
transition to the true vacuum state produces a non-zero
cosmological constant and a de Sitter phase of inflation.

We shall describe the phase transition to fermion condensates
using a non-relativistic toy model. The Hamiltonian takes the BCS
form with ${\bf k}=-{\bf k}$~\cite{Taylor}: \be
\cal{H}\rm=\sum_{k's',ks}{\cal E}_k c_{ks}^\dagger c_{ks}
-\frac{1}{2}\sum_{k,k',s,s'}V_{kk'}c_{k's'}^\dagger
c_{-ks}^\dagger c_{-k's'}c_{ks} \ee

We perform the transformation to new operators
\be
b_k=u_kc_k-v_kc^\dagger_{-k},\quad b_{-k}=u_kc_{-k}+v_kc^\dagger_k
\ee
where the b's satisfy anti-commutation relations and
$u^2_k+v_k^2=1$. The fermion number operators are $n_k=b_k^\dagger
b_k$ and $n_{-k}=b_{-k}^\dagger b_{-k}$.

We must now determine the ground state (vacuum) and set the
occupation numbers $n_k$ and $n_{-k}$ equal to zero. We need to
determine the energy gap $\Delta$ produced by the gap in the
vacuum energy in the phase transition to the fermion condensates.
The condensates are bound states due to the weak gravitational
interaction generated by the exchange of gravitons between
fermions and the screened attractive force. This will give
\[
\Delta=\sqrt{\Lambda}
\]

We can minimize the Hamiltonian energy by diagonalizing $\cal{H}\rm$,
giving the condition \be \label{zerodiag} {\cal
E}_k\biggl(\frac{1}{4}-x_k^2\biggr)^{1/2} +x_k
\sum_{k'}V_{kk'}\biggl(\frac{1}{4}-x_{k'}^{2}\biggr)^{1/2}=0 \ee where
$u_k=(\frac{1}{2}-x_k)^{1/2}$ and $v_k=(\frac{1}{2}+x_k)^{1/2}$,
and $V_{kk'}$ is the interaction matrix associated with the
exchange of gravitons and the screened attractive fermion force.

We define the quantity \be \label{Delta}
\Delta_k=\sum_{k'}V_{kk'}\biggl(\frac{1}{4}-x_{k'}^{2}\biggr)^{1/2}
\ee Then (\ref{zerodiag}) yields \be\label{xequation} x_k=\pm
\frac{{\cal E}}{2({\cal E}^2+\Delta_k^2)^{1/2}} \ee By
substituting this into (\ref{Delta}), we obtain the integral
equation for $\Delta_k$: \be
\Delta_k=\frac{1}{2}\sum_{k'}V_{kk'}\frac{\Delta_{k'}}{({\cal
E}^2_{k'} +\Delta_{k'}^2)^{1/2}} \ee

We assume the simple model for the interaction matrix: \be
V_{kk'}=V\quad {\rm if}\,\, \vert {\cal E}_k\vert < \omega_D,
$$ $$
V_{kk'}=0\quad {\rm otherwise} \ee where $\omega_D$ ($\hbar=1$) is
the Debye energy and $V$ is a constant. Choosing the minus sign in
(\ref{xequation}), we obtain for the energy gap \be
\Delta_k=\frac{1}{2}VD\int^{\omega_D}_{-\omega_D}d{\cal
E}\frac{\Delta}{({\cal E}^2+\Delta^2)^{1/2}} \ee The solution to
this equation is \be\label{sinhequation}
\Delta=\sqrt{\Lambda}=\frac{\omega_D}{\sinh[1/VD]} \ee where $D$
is the fermion density of states defined by the Fermi sphere for
$N$ fermions by \be n_f=\int_{k_0}^{k_f}d^3kD(k) \ee

The physical density of fermion states behaves for an expanding
universe as \be D(\omega_k)\sim a^3(t) \ee so that as the universe
inflates and $a(t)\rightarrow\infty$, we have
$D(\omega)\rightarrow \infty$. However, $VD$ is independent of the
cosmic scale $a(t)$.

In the early universe there is an initial phase in which spacetime
is flat (Minkowski) and the fermions do not interact
gravitationally. This phase is unstable to gravitational
interactions between fermions. There is a phase transition to an
inflating de Sitter vacuum, in which we have a {\it strong
coupling} limit $VD\sim 1$ and \be \Delta_i=\sqrt{\Lambda}\sim
\omega_D. \ee In this phase \be \sqrt{\Lambda}\sim \omega_D \sim
M_{\rm PL} \ee where $M_{\rm PL}$ is the Planck mass.  As the
universe expands exponentially, a weak coupling limit develops
when \be VD\ll 1 \ee which from (\ref{sinhequation}) leads to an
exponential suppression of the cosmological constant\be
\label{weaklimit}
\Delta_f=\sqrt{\Lambda}=2\omega_D\exp\biggl(-\frac{1}{VD}\biggr)
\ee

The weak coupling limit (\ref{weaklimit}) can be interpreted as a
weakening of the correlation between the fermions associated with
a decay of the vacuum energy into pairs of particles at the end of
inflation. As $\Lambda$ tends to zero the universe enters the
radiation dominated phase of an FRW model. We see that the
condensate phase can generate enough inflation initially and then
produce an exponential suppression of the cosmological constant,
leading to a vanishingly small value of $\Lambda$ in the present
universe.

The condensation energy for the weak coupling limit $VD\ll 1$ is
given by \be {\cal E}_{\rm cond}\sim
-2\omega_D^2D\exp\biggl(-\frac{2}{VD}\biggr)\sim
-\frac{1}{2}D\Lambda \ee As the universe expands from its initial
inflationary period, the number of fermions that is affected by
the attractive gravitational and screened interaction is a small
fraction of the total number of fermions in the universe. We note
that $\exp[-2/(VD)]$ has an essential singularity at $V=0$, which
means that while the function and its derivatives vanish as
$V\rightarrow +0$, they all become infinite as $V\rightarrow -0$.
This means that we cannot calculate ${\cal E}_{\rm cond}$ by using
perturbation theory.

In order for our mechanism to produce a suppression of the
cosmological constant, we must have a large enough density of
fermions $D$ in the de Sitter phase of inflation. Inflation
produces enormous numbers of massless, minimally coupled scalar
condensates $\phi=\langle\bar\psi\psi\rangle$. The conformal
invariance of free Dirac theory implies that there can be no
comparable, direct production of fermions. However, it is possible
to produce fermions during inflation by allowing them to interact
with a massless, minimally coupled scalar or fermion condensate
$\langle\bar\psi\psi\rangle$~\cite{Prokopec}. The physical
interpretation is that inflation alters the constraint of energy
conservation to permit the spontaneous appearance of a condensate
and a fermion-anti-fermion pair, and the fermions do not recombine
to make virtual pairs. This mechanism could produce a large number
of fermions in the de Sitter space and allow for a non-zero
fermion density $D$ when the fermion occupation numbers $n_k$ and
$n_{(-k)}$ are zero.

\section{Vacuum Instability in the de Sitter Phase}

In the previous section, we provided general arguments for the
vacuum instability which drives the universe into a de-Sitter
inflationary phase. We also demonstrated that during inflation the
weakening of correlations between the fermions will lead to an
exponential suppression of the cosmological constant. We now
present a relativistic model which shows that the perturbative
(false) vacuum in de-Sitter space is unstable in the presence of
gravitational interactions to a non-perturbative true vacuum
state.  Instead, the perturbative graviton interaction between
fermion pairs drives the fermions into non-perturbative condensate
states \footnote { The vacuum energy can transmute into massive
degrees of freedom, this possibility is currently under
investigation by the authors}. In the case of inflation, these
states are the Goldstone bosons corresponding to the broken de
Sitter space-time symmetry, which commences with a large and then
exponentially suppressed cosmological constant.

Let us illustrate this mechanism with a model which effectively
models the formation of Cooper-pairs in a relativistic
gravitational context. The important point is that our Lagrangian
(\ref{action}) will get modified by an interaction Lagrangian
which takes into account graviton exchange between pairs of
fermions.  Consider the following theory with a massless, free
fermion coupled to gravity. For illustration, we consider the
modified version of eq (\ref{action}): \be \cal{L}\rm = \sqrt{g}[R
+ \sum_{k=1}^{N}\bar{\psi}D_{a}\gamma^{a}\psi \;\; +
\sum_{k=1}^{N} \frac{G}{2N}(\bar{\psi}\psi\bar{\psi}\psi)]
\label{NJL} \ee where $D_a$ is the covariant derivative with
respect to the local spin connection, $G$ is the gravitational
coupling constant, $N$ is the number of fermion species, and the
third term is a four-fermion interaction, which at the Fermi
surface describes the relevant graviton interaction between pairs
of fermions on the Fermi surface.

It is well-known that the physics described by (\ref{NJL}) is
equivalent to the following Lagrangian \be \cal{L}\rm = \sqrt{g}[R
+ \bar{\psi}D_{a}\gamma^{a}\psi\;\; + \bar{\psi}\phi\psi -
\frac{N}{2G}\phi^{2}] \ee where $\phi=<\bar{\psi}\psi>$ is the
condensate which forms from fermion-graviton interactions.

We will consider graviton exchange between pairs of fermions by
expanding about Minkowski spacetime \be
g_{\mu\nu}=\eta_{\mu\nu}+h_{\mu\nu}+O(h^2)\ee where
$\eta_{\mu\nu}$ is the background Minkowski space metric. For a
Dirac fermion, we obtain for a momentum cutoff $K_c$ in the weak
curvature limit: \be i\gamma^\mu
p_\mu+\sqrt{\Lambda_0}+\Sigma(p,\sqrt{\Lambda},G,K_c)=0\ee for
$i\gamma^\mu p_\mu+\sqrt{\Lambda}=0$\footnote{A de Sitter space
solution to the non-perturbative gap equation has been obtained
for a large $N$ expansion by Inagaki et al.~\cite{Inagaki}.}.

We have for a zero bare cosmological constant, $\Lambda_0=0$: \be
\sqrt{\Lambda}=\Sigma \ee For gravity for the lowest-order loop we
obtain\be \sqrt{\Lambda} =G_0\sqrt{\Lambda}F(\sqrt{\Lambda},K_c)
\ee where $F(\sqrt{\Lambda},K_c)$ is the result of the momentum
integration of the Feynman fermion propagators and the cutoff is
$K_c=M_{\rm PL}$, where $M_{\rm PL}$ is the Planck mass. This has
two solutions: either $\sqrt{\Lambda}$ is zero or \be
\frac{1}{G_0}=F(\sqrt{\Lambda},K_c) \label{gap} \ee The first
solution is the trivial perturbative solution, while the second,
nontrivial non-perturbative solution determines $\sqrt{\Lambda}$
in terms of the bare gravitational coupling constant $G_0$ and the
cutoff $K_c$. The nontrivial solution corresponds to the
superfluid condensate state which is the true vacuum state of the
system, while the trivial solution corresponds to the normal
(false) vacuum state i.e. not the true vacuum state.

The gap equation (\ref{gap}) asymptotically has an exponential
dependence on the physical density of states.  We have that in an
FRW de Sitter background filled with fermions the energy gap will
take on the following form, $\Delta=\sqrt{\Lambda}$, separating
the two phases. Physically this means that the gap corresponds to
the binding energy necessary to form the condensate. The
difference between the original vacuum energy and the final vacuum
energy is the rest mass of the condensate.

\section{Conclusions}

We have shown that the vacuum energy in the early universe can
become unstable as the attractive fermion-gravitational force and
a screened attractive force between positively and
negatively charged fermions produces condensates through a phase
transition at a critical temperature $T < T_c$. An initial phase
of Minkowski flat spacetime with a zero cosmological constant is
unstable through a transition to a de Sitter vacuum with an onset
of inflation, caused by a non-perturbative vacuum with a large
vacuum energy gap $\Delta=\sqrt{\Lambda}$. When the universe
ceases to inflate an exponential suppression relaxes the
cosmological constant to a small or zero value in the present
universe. We described this scenario by analogy with the BCS
mechanism associated with the formation of Cooper pairs of
fermions by means of the exchange of gravitons.

The non-perturbative mechanism can explain how an initially large
vacuum energy (cosmological constant) can be suppressed by a phase
transition to a superfluid state of the early universe as the
universe expands, leading to a ``graceful'' exit for inflation.

The non-relativistic toy model we have used to describe the
scenario can be extended to a relativistic QFT model of the
formation of a vacuum energy gap for fermion condensates in a de
Sitter spacetime background.

This scenario explains why a naive {\it perturbative} calculation
of the vacuum energy leads to a nonsensical answer. In contrast to
the ground state or vacuum of QED or the standard model, the
vacuum associated with gravity is unstable and the instability can
only be described by non-perturbative physics.

We note that our model differs from a fundamental scalar field
with a tuned potential because the gravitational vacuum `knows'
about the composite nature of the condensate unlike for
fundamental scalar fields. Our scenario precludes the existence of
elementary scalar field particles such as the standard Higgs
particle. The Higgs particle is pictured as a composite of
fermion-antifermion pairs. The same holds true for the graviton
which is described in our picture as a composite condensate of
four fermions forming a spin-2 graviton~\cite{Wetterich}.

In future work a more detailed investigation will be carried out
of the properties of the gravitational self-energy of the fermions
and the role played by the energy gap in early universe cosmology.  Since the energy scales during inflation is in the regime of the deconfining phase of QCD it is of interest to see how the gap equation is modified in the presence of free quarks at finite density \cite{Alford}. It is also important to understand how this mechanism fares with
other contributions to the vacuum energy such as composite bosonic
degrees of freedom.

\section{Acknowledgements}

The work of SHSA is supported by the US DOE under grant DE-AC03-76SF00515.  JM acknowledges
the support of the Natural Sciences and
Engineering Research Council of Canada. We especially thank BJ Bjorken, Robert
Brandenberger, Justin Khoury, Michael Peskin, M.M Sheikh-Jabbari and Lenny Susskind for helpful discussions.  We also thank Niayesh Afshordi, Simeon
Hellerman, Ronald Mallett, Paul Steinhardt and Richard Woodard
for discussions. \vskip 0.5 true in

\end{document}